\documentclass{PoS}
\newcommand{\lyxdot}{.}
\usepackage{float}
\usepackage{amsmath}
\usepackage{graphicx}
\usepackage{setspace}
\usepackage{esint}
\usepackage{subfigure}

\title{
 \begin{picture}(0,0)(0,0)%
   \put(350,75){\makebox(0,0)[l]{\textnormal{\normalsize UTHEP-665}}}%
   \put(350,65){\makebox(0,0)[l]{\textnormal{\normalsize CERN-PH-TH-2014-258}}}%
   \put(350,55){\makebox(0,0)[l]{\textnormal{\normalsize KEK-CP-317}}}%
   \end{picture}%
	Scaling study of an improved fermion action on quenched lattices
}

\ShortTitle{Scaling study of an improved fermion action on quenched lattices}

\author{\speaker{Yong-Gwi Cho} \\
  Graduate School of Pure and Applied Sciences, University of Tsukuba,\\
  Tsukuba, Ibaraki 305-8571.\\
  E-mail: \email{cho@ccs.tsukuba.ac.jp}}

\author{{Shoji Hashimoto, Jun-Ichi Noaki, Takashi Kaneko}\\
  High Energy Accelerator Research Organization (KEK) and\\
  School of High Energy Accelerator Science, The Graduate University
  for Advanced Studies,\\
  Tsukuba 305-0801.
}

\author{Andreas J\"uttner, Justus Tsang\\
School of physics and astronomy, University of Southampton,\\
Highfield, Southampton SO17 1BJ, UK
}  

\author{{Marina Marinkovic}\\
School of Physics and Astronomy, University of Southampton \\
Highfield, Southampton SO17 1BJ, UK\\
CERN, Physics Department, 1211 Geneva 23, Switzerland\\
}

\abstract{We present scaling studies for heavy-quark observables calculated with an $O(a^2)$-improved fermion action on tree-level Symanzik improved gauge configurations. Lattices of $1/a = $ 2.0-3.8 GeV with an equal physical volume 1.6 fm are used. The results are compared with the standard domain-wall and naive Wilson fermions.}

\FullConference{The 32nd International Symposium on Lattice Field Theory,\\
		23-28 June, 2014\\
		Columbia University New York, NY}
		
\begin{document}

\section{Introduction}
Precise determination of the Standard Model (SM) parameters such as the CKM matrix elements $V_{ub},V_{cb}$ is a necessary step in the search for new physics through precision measurements.
In many processes that are sensitive to new physics, charm and bottom quarks are included in the hadronic initial and/or final state.
To estimate the hadronic contribution, lattice QCD plays a crucial role, but the heavy quark discretization effect can be overwhelming for $m_qa$ on currently available gauge configurations.
While the effective field theory approach for heavy quarks on the lattice is applicable for the calculation of these quantities, a brute-force approach of taking $a$ as small as possible with conventional quark actions is also powerful as there is no need of additional matching of parameters. 
This may be combined with the Symanzik improvement of the lattice fermion action to eliminate leading discretization effects.
In this work we investigate some choices of the lattice fermion action to be used for heavy valence quarks, focusing on their discretization effect and continuum scaling.
In particular, we are interested in an improved action based on the Brillouin fermion \cite{Durr:2010ch}.
In this initial study, we mainly consider the charm quark mass region, and leave an extension towards heavier masses for future study. 
The quantities to be studied are the energy-momentum dispersion relation and hyperfine splitting of the heavy-heavy mesons.
For numerical tests, we have generated quenched gauge configurations that have a roughly matched physical volume 
(at about 1.6~fm) and cover a range of lattice spacings between $1/a$ = 2.0 and 3.8~GeV. 
Since these lattices do not contain sea quarks and have a small physical volume, we do not expect precise agreement with the corresponding experimental data for the charm quark,  but rather we are interested in their scaling towards the  continuum limit.
Here we update the report \cite{Cho:2013yha} by adding a finer gauge configurations.

\section{Formalism and simulation details}
We focus on the discretization effects of the Brillouin fermion \cite{Durr:2010ch} and propose its improved version according to the Symanzik  improvement program \cite{Cho:2013yha}.
Here let us briefly describe the improved action.
The improved Dirac-operator is given by 
\begin{equation}
D^{imp}=\underset{\mu}{\sum}\gamma_{\mu}(1-\frac{a^{2}}{12}\triangle^{bri})\nabla_{\mu}^{iso}(1-\frac{a^{2}}{12}\triangle^{bri})+c_{imp}a^{3}(\triangle^{bri})^{2},
\label{eq:imp_bri_action}
\end{equation}
where the operators $\nabla^{iso}_{\mu}$ and $\triangle^{bri}$ are the isotropic derivative and the Brillouin laplacian operator, respectively.
These operators can be defined as 
\begin{align}
\nabla^{iso}_{\mu}\left(n,m\right) & =\rho_{1}\left[\delta_{n+\hat{\mu},m}-\delta_{n-\hat{\mu},m}\right]+\rho_{2}\underset{\nu}{\sum}\left[\delta_{n+\hat{\mu}+\hat{\nu},m}-\delta_{n-\hat{\mu}+\hat{\nu},m}\right]\nonumber \\
&+\rho_{3}\underset{\nu,\rho}{\sum}\left[\delta_{n+\hat{\mu}+\hat{\nu}+\hat{\rho},m}-\delta_{n-\hat{\mu}+\hat{\nu}+\hat{\rho},m}\right]
+\rho_{4}\underset{\nu,\rho,\sigma}{\sum}\left[\delta_{n+\hat{\mu}+\hat{\nu}+\hat{\rho}+\hat{\sigma},m}-\delta_{n-\hat{\mu}+\hat{\nu}+\hat{\rho}+\hat{\sigma},m}\right],
\end{align}

\begin{align}
\triangle^{bri}\left(n,m\right) & =\lambda_{0}\delta_{n,m}+\lambda_{1}\underset{\mu}{\sum}\delta_{n+\hat{\mu},m}+\lambda_{2}\underset{\mu,\nu}{\sum}\delta_{n+\hat{\mu}+\hat{\nu},m}\nonumber \\
&+\lambda_{3}\underset{\mu,\nu,\rho}{\sum}\delta_{n+\hat{\mu}+\hat{\nu}+\hat{\rho},m}
+\lambda_{4}\underset{\mu,\nu,\rho,\sigma}{\sum}\delta_{n+\hat{\mu}+\hat{\nu}+\hat{\rho}+\hat{\sigma},m},
\end{align}
with $\left(\rho_{1},\rho_{2},\rho_{3},\rho_{4}\right)=\frac{1}{432}\left(64,16,4,1\right)$ and
$\left(\lambda_{0},\lambda_{1},\lambda_{2},\lambda_{3},\lambda_{4}\right)=\frac{1}{128}\left(240,-8,-4,-2,-1\right)$.
The summations over $\mu,\nu,\rho,\sigma=\pm\hat{1},\pm\hat{2},\pm\hat{3},\pm\hat{4}$ are taken in different  directions, i.e. $\mu\neq\nu\neq\rho\neq\sigma$.
The simplest way of constructing a gauge covariant operator is to sum over all shortest paths connecting neighboring sites. 
$D^{imp}$ has no $O(a)$ and $O(a^2)$ errors at tree-level and its leading discretization error is of $O(a^3)$. 
$c_{imp}$ is a free parameter to be set.

In this study, we also use the stout link smearing with which one expects that the radiative corrections of $O(\alpha_s a)$ and $O(\alpha_s a^2)$ are not substantial.
In this work, we examine the scaling property of some heavy-heavy meson observables, such as the speed-of-light and hyperfine splitting.
For the gauge action we employ the tree-level improved Symanzik action.
For the ease of comparison with the Wilson or the domain-wall operator, we set $c_{imp}=1/8$ as a first choice.
We are planning to tune $c_{imp}$ beyond this tree-level value in order to realize more continuum-like Dirac spectrum.
As listed in Table \ref{table1}, three quenched gauge ensembles with lattice spacing between $1/a=$ 2.0 and 3.8 GeV have been generated keeping the physical box size roughly fixed to $L\approx1.6$ fm.
The lattice spacing is determined using the Wilson-flow scale $w_0$ \cite{Borsanyi:2012zs}. 
For the detail of the gauge ensembles, please see a similar scaling study using same set of ensembles \cite{Tobi:2014lat}.
We compare the scaling properties among the improved Brillouin Dirac operator $D^{imp}$, the naive Wilson Dirac operator and the domain-wall Dirac operator. 
The latter two have $O(a)$ and $O(a^2)$ leading discretization errors, respectively.
For the implementation of the domain-wall fermions, we take the M\"{o}bius domain-wall fermion discretization with three steps of stout smearing, as used in the recent works of the JLQCD collaboration \cite{Noaki:2014}.
We set the size of the fifth dimension $L_{s}=8$ and the domain-wall height $M_{0}=-1$.
The violation of the Ginsparg-Wilson relation, as measured by the residual mass $m_{res}$, gives $m_{res} \lesssim O(0.1)\;\rm{MeV}$ or smaller.
The code used in this study is developed as a part of the IroIro++ package \cite{Cossu:2013ola}.

For the comparison of different types of Dirac operators, we roughly tune the bare quark mass to give pseudo-scalar meson masses of  $m_{ps}=1.0$, $1.5$, $2.0$, $2.5$, $3.0$ and $3.5$ [GeV].
In the calculation of the quark propagator, we use smeared sources with a function $e^{-\alpha\left|x-y\right|}$.
Our main data set is obtained for $m_{PS}=3.0$ GeV, where the calculation is repeated with 4 source positions to improve the signal.
For other $m_{PS}$, we use a single source location.
We obtain the results at our target values of $m_{PS}$ by interpolating the fitted values of the pseudo-scalar meson masses.
\begin{center}
\begin{table}[t]
\begin{centering}
\begin{tabular}{ccccc}
\hline 
$L/a$ & $\beta$ & $N_{conf}$ & $a^{-1}${[}GeV{]}\tabularnewline
\hline 
16 & 4.41 & 100 & 2.00(07) \tabularnewline
\hline 
24 & 4.66 & 100 & 2.81(09) \tabularnewline
\hline 
32 & 4.89 & 100 & 3.80(12) \tabularnewline
\hline 
\end{tabular}
\par\end{centering}
\caption{Parameters, numbers of statistics and the measured values of the lattice spacings of the gauge configurations used in this study. The lattice volume is fixed to $L\approx1.6$ fm.}
\label{table1}
\end{table}
\par\end{center}

\section{Speed-of-light for pseudo-scalar meson}
An effective speed-of-light $c_{\mathrm{eff}}(p)$ is defined as
\begin{equation}
c_{\mathrm{eff}}^{2}\left(p\right)=\frac{E^{2}(\overrightarrow{p})-E^{2}(\overrightarrow{0})}{\overrightarrow{p}^{2}},
\label{eq:speed-of-light}
\end{equation}
which should be unity in the continuum theory. 
This provides a basic test of the Lorentz symmetry violating discretization effects, which becomes more significant for heavy quarks since $m_{q}a$ increases.
We calculate the two-point correlation functions for various momenta as 
$C\left(t,\overrightarrow{p}\right) =\underset{x}{\sum}\left\langle M\left(\overrightarrow{x},t\right)\overline{M}(\overrightarrow{0},0)\right\rangle e^{i\overrightarrow{p}\overrightarrow{x}},$
with the meson operator $M\left(\overrightarrow{x},t\right)=\bar\psi(x)\gamma_5\psi$ for the pseudo-scalar meson.
We obtain the energy from the exponential fall-off of the correlators and extract the speed-of-light for each value of $\overrightarrow{p}$.
The speed-of-light calculated for the mass of $m_{ps}=3.0$ GeV at the coarsest lattice ($1/a=2.0$ GeV) is shown in Figure \ref{spl_v16}.
In this figure, the data for the Wilson fermion (red squares), improved Brillouin fermion (blue triangles) and domain-wall fermion (magenta diamonds) are plotted as a function of normalized momentum $\left|\overrightarrow{p}^2\right|(L/2\pi)^2$.
As we clearly see, the data for the improved Brillouin action is consistent with unity within statistical errors, while the data for the Wilson fermion and the domain-wall fermion deviate from the Lorentz-symmetric value by about 20-30$\%$. This is understood already at tree-level, and clearly demonstrates the advantage of the Brillouin-type operator as already reported in \cite{Durr:2012dw} for an unimproved version.

\begin{center}
\begin{figure}[!htb]
\begin{centering}

\includegraphics[scale=0.4]{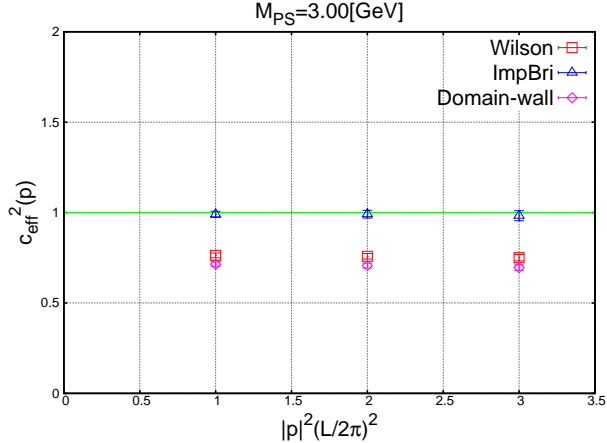}

\par\end{centering}
\vspace{10pt}
\caption{Effective speed-of-light calculated on the coarsest lattice $a^{-1}=1.973$ GeV for $m_{ps}=3.0$GeV. The three colors represent the three different discretizations of the Dirac operators.
\label{spl_v16}}
\end{figure}

\par\end{center}

We show the scaling of the speed-of-light against $a^{2}$ for the meson mass of $m_{ps}=3.0$ GeV in Figure \ref{scaling_spl}. 
One can clearly observe that the scaling of the improved action is excellent.
We do not see any significant deviations from unity for the ensembles with $1/a=2.0-3.8$ GeV.
For all momenta which are shown in Figure \ref{spl_v16} and lattices, the speed-of-light of the improved action is unity within statistical errors.

To investigate scaling properties, we try to model the discretization effect using the data at $|p|^2(L/2\pi)^2=1$.
One may employ an ansatz $c^2(a)=1+c_1a+c_2a^2$, since one expects that the speed-of-light converges to unity in the continuum limit.
For the domain-wall fermions, an ansatz $c^2(a)=1+c_2a^2+c_4a^4$ should be used instead, because the $O(a)$ and $O(a^3)$ terms are forbidden by chiral symmetry. 
Strictly speaking, there might be $O(a)$ and $O(a^3)$ discretization effects because of non-zero $m_{res}$, but given the tiny $m_{res}$ such effects are negligible.

For the Wilson fermion, we obtain $c_1=0.18(9)$ and $c_2=-26(1)$ with $\chi^2/d.o.f=0.048$.
For improved Brillouin fermion, $c_1=0.5(4)$ and $c_2=-5(5)$ with $\chi^2/d.o.f=0.826$.
The large error for $c_{1}$ and $c_{2}$ implies that the lattice data are consistent with vanishing $O(a)$ and $O(a^{2})$ effects.
If one assumes an ansatz $c^2(a)=1+c_3a^3+c_4a^4$ for the improved action, we obtain $c_3=166(74)$ and $c_4=-1777(778)$ with $\chi^2/d.o.f=0.311$.
Though the error is still large, their size is consistent with nominal values $(1 \rm{GeV})^{3}$ and $(1 \rm{GeV})^{4}$, as expected.
Thus we may conclude that $O(\alpha_sa)$ and $O(\alpha_sa^2)$ discretization effects are negligible for the improved action.
For the domain-wall fermion, we found $c_2=-18(2)$ and $c_4=-1033(271)$ with $\chi^2/d.o.f=0.512$. The size of these coefficients is again consistent with 1 GeV and   $(1\rm{GeV})^{2}$.


\begin{center}
\begin{figure}[!htb]
\begin{centering}

\includegraphics[scale=0.3]{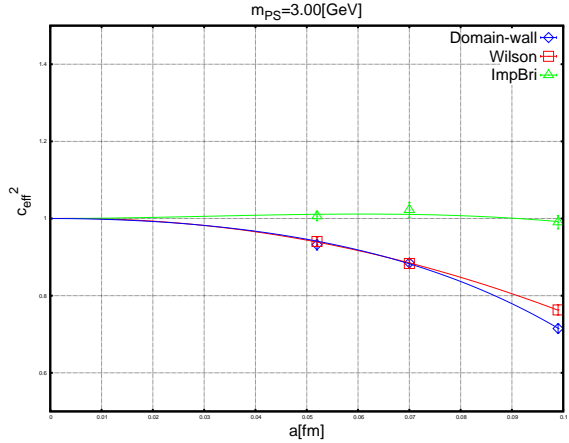}
\par\end{centering}

\vspace{10pt}
\caption{The scaling of the speed-of-light against $a^2$  at $m_{ps}=3.0$ GeV, where the used ansatz is $f(a)=1+c_1a+c_2a^2$ for the Wilson fermion, $f(a)=1+c_2a^2+c_3a^3$ for the improved action and $f(a)=1+c_2a^2+c_4a^4$ for  the domain-wall fermion.}
\label{scaling_spl}
\end{figure}
\par\end{center}

\section{Hyperfine splitting}

Hyperfine splitting $m_{V}-m_{PS}$ is also useful to test the discretization effect for heavy quarks and has been investigated with various lattice actions.
It is well-known (see, e.g. \cite{Okamoto:2001jb,Stewart:2000ev}), that the charmonium hyperfine splitting strongly depends on the choice of the Dirac operator, which means that the hyperfine splitting is very sensitive to the discretization error. 
We show the scaling of the hyperfine splitting in Figure \ref{scaling_hyp_a^2}.
The Wilson fermion may contain an $O\left(a\right)$ error whilst the leading order errors of the domain-wall fermions are $O\left(a^{2}\right)$.
For the improved Brillouin operator, we observe that the scaling is very mild against $a^{2}$.
Theoretically, the leading discretization effect is $O\left(a^{3}\right)$ as well as $O\left(\alpha_sa\right)$.

Here again, we examine the fitting with an ansatz $f(a)=c_0+c_1a+c_2a^2$.
Note that the experimental value of the charmonium hyperfine splitting is known, but since the lattice data receive the systematics effect due to quenching we can not reproduce the experimental value. 
Thus we take a common constant $c_0$ for all three Dirac operators and fit all data simultaneously with the ansatz.
We obtain $c_0=0.0787(1)$ and $c_2=-4.1(2)$ for the domain-wall fermion. 
For the Wilson fermion, $c_1=-1.2(4)$ and $c_2=4.9(3)$ are observed.
For the improved action, we find $c_1=-0.38(5)$ and $c_2=2.2(5)$.
The fit is statistically reasonable $\chi^2/d.o.f=0.16$.
We found smaller coefficients $c_1$ and $c_2$ for the improved action than the $c_1$ of the Wilson fermion and $c_2$ of the domain-wall fermion.
We also attempted to fit the data of the domain-wall fermion by the ansatz $f(a)=c_0+c_2a^2+c_4a^4$, but the error for $c_4$ was quite large.
Thus instead we restricted the ansatz to $f(a)=c_0+c_2a^2$ only for the fit of the domain-wall data.

If one assumes that $O(\alpha_sa)$ is small for the improved action, an ansatz $f(a)=c_0+c_2a^2+c_3a^3$ might be adopted, from which we found  $c_2=-9(1)$ and $c_3=72(10)$ with $\chi^2/d.o.f=0.28$.
However, then $c_2$ is unreasonably large and it suggests that $f(a)=c_0+c_2a^2+c_3a^3$ is not an appropriate ansatz.
Thus, we can not conclude that $O(\alpha_sa)$ is negligible for the improved action. 
Here it seems natural to conclude that $O(\alpha_sa)$ is still visible, but its effect is smaller than that of the other formalisms.

\begin{center}
\begin{figure}[!htb]
\begin{centering}

\begin{centering}
\includegraphics[scale=0.3]{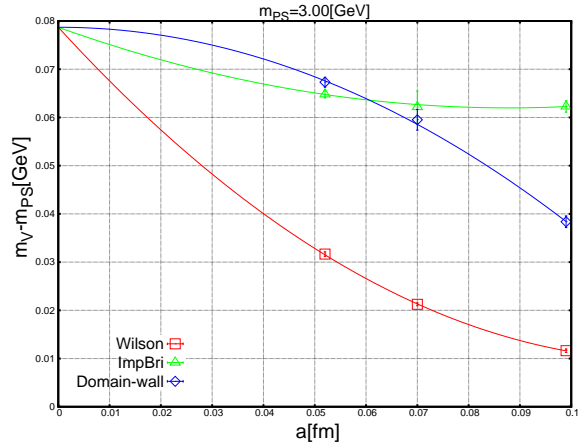}
\par\end{centering}

\par\end{centering}
\vspace{10pt}
\caption{Scaling for the hyperfine splitting against $a^2$, where the used ansatz is $f(a)=1+c_1a+c_2a^2$ for the Wilson fermion and the improved action, $f(a)=1+c_2a^2$ for the domain-wall fermion.}
\label{scaling_hyp_a^2}
\end{figure}

\par\end{center}
\section{Summary}
We have shown non-perturbative scaling studies of the improved Brillouin fermion action compared with two other fermion actions.
From the scaling of the speed-of-light, we see that scaling of the improved action is excellent, always obeying the continuum dispersion relation.
From the scaling of the hyperfine splitting, we also observe good scaling of the improved action.
We note that our improvement is still at tree-level and one-loop level improvement or non-perturbative tuning of the coefficients is desirable.

The computational cost to solve the heavy quark propagator for the improved Brillouin fermion is only about 10 times higher than that for the Wilson fermion, though a single application of the Dirac operator is 
more than 100 times costly. 
This is because the condition number for the Brillouin operator is much lower than that for the Wilson fermion.
Thus, given its excellent scaling property, the improved Brillouin fermion is a promising formulation for heavy valence quarks. 
On the Blue Gene/Q machine we are using for this work, the Wilson-Dirac operator is highly 
optimized, and more work is necessary to optimize the improved Brillouin operator to that level. 

\vspace*{0.8cm}

We would like to thank Stephan D\"{u}rr for detailed discussions and offering codes.
Numerical simulations are performed on the IBM System Blue Gene Solution at High Energy
Accelerator Research Organization (KEK) under a support of its Large Scale Simulation Program
(No. 13/14-04). This work is supported in part by the Grant-in-Aid of the Japanese Ministry of
Education (No. 26247043) and the SPIRE (Strategic Program for Innovative Research) Field5 project.
The research leading to these results has received funding from the European Research Council under the European Union's Seventh Framework Programme (FP7/2007-2013) / ERC Grant agreement 279757.

\bibliographystyle{unsrt}
\bibliography{./brillouin_hq.bib}

\end{document}